# Efficient use of single molecule time traces
# to resolve kinetic rates, models and uncertainties.

Sonja Schmid[1,2], Thorsten Hugel[1]

[1] *Institute of Physical Chemistry, University of Freiburg, Germany*
[2] *Kavli Institute for Nanoscience, Delft University of Technology, Netherlands*

**Abstract**

Single molecule time traces reveal the time evolution of unsynchronized kinetic systems. Especially single molecule Förster resonance energy transfer (smFRET) provides access to enzymatically important timescales, combined with molecular distance resolution and minimal interference with the sample. Yet the kinetic analysis of smFRET time traces is complicated by experimental shortcomings - such as photo-bleaching and noise.

Here we recapitulate the fundamental limits of single molecule fluorescence that render the classic, dwell-time based kinetic analysis unsuitable. In contrast, our Single Molecule Analysis of Complex Kinetic Sequences (SMACKS) considers *every* data point and combines the information of many short traces in one global kinetic rate model.

We demonstrate the potential of SMACKS by resolving the small kinetic effects caused by different ionic strengths in the chaperone protein Hsp90. These results show an unexpected interrelation between conformational dynamics and ATPase activity in Hsp90.

## INTRODUCTION

Single molecule time traces are particularly suited to investigate kinetic and thermodynamic questions in molecular machines, such as proteins. In fact, it is the unique feature of single molecule time traces, to reveal the time evolution of one molecule through individual kinetic states - notably in real time and at steady-state, without the need for external synchronization. This allows one to explore the energy landscape or to uncover the molecular driving force powering a protein's function.

Single molecule Förster resonance energy transfer (smFRET) is a popular method to study protein folding as well as native dynamics[1-6]. However, the experimental detection of single molecule fluorescence time traces is complicated by the antagonistic relation between the three key numbers illustrated in Figure 1: signal-to-noise ratio (SNR), time resolution and observation time. An excellent SNR - although itself desirable - requires a relatively high excitation power. This comes with faster photo-bleaching and, thus, a reduced observation time for a given fluorophore. Likewise, high time resolution - i.e. a fast sampling rate - needs even higher excitation powers to reach an equivalent SNR at shorter exposure times. At the core of this vicious circle lies the finite number of photons an individual fluorophore can emit - typically a few million photons[7,8] - before it undergoes irreversible photo-bleaching.

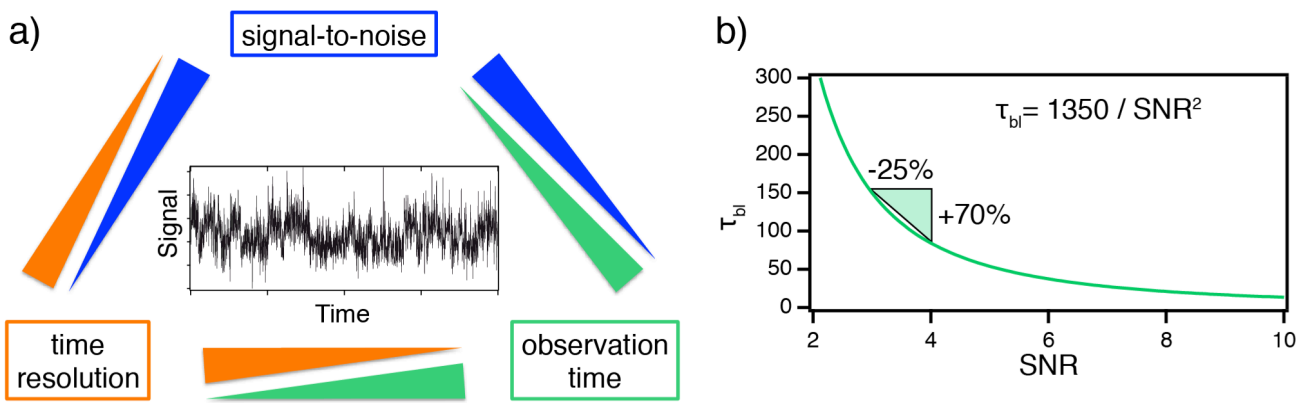

**Figure 1:** (a) In smFRET experiments there is an antagonistic relation between signal-to-noise ratio (SNR), time resolution and observation time. Increasing one of the three has a negative effect on the remaining two. (b) The empirical relation of observation time (limited by the time constant of photo-bleaching, $\tau_{bl}$) and SNR derived from the experiment (see section Theory). A SNR of about 4 comes with $\tau_{bl} = 90$ frames in typical alternating laser excitation (ALEX) experiments, yielding the empirical $const = 1350$. As indicated, a reduction of 25% in SNR may result in as much as 70% longer observation time.

Within the regime of realistic TIRF experiments, a simple relation between observation time $\tau_{bl}$ and SNR can be derived (see Theory), namely:

$$\tau_{bl} = \frac{const.}{SNR^2} \qquad (1)$$

It holds regardless of the specific time resolution, i.e. sampling rate, of the experiment, and helps to decide how to best spend the precious photons in an experiment. In contrast to pure distance determination experiments, where a high signal-to-noise ratio is the only goal, *kinetic* analysis also requires a large enough observation time for the detection of possibly complex dynamics (e.g. kinetically heterogeneous behavior). It can be shown in this case that sacrificing SNR to a certain extent yields a disproportionately high increase in observation time, thus augmenting the total gain in kinetic information.

Nevertheless, the maximum detection bandwidth achieved in single molecule fluorescence time traces is still remarkably low - even among single molecule techniques. This is problematic for any dwell-time based analysis, which has been the recognized standard in single molecule kinetics for decades[6,9]. While it may have been adequate for long patch clamp trajectories, it is clearly unfit for shorter fluorescence traces, as well as more and more complex dynamics[10]. In fact, it is biased towards short dwells, which causes a systematic overestimation of all rates. And more severely, it can lead to even qualitatively wrong interpretations. Additional complication arises from experimental noise and signal variations between individual molecules (i.e. different intensities of individual fluorophores).

To cope with these shortcomings and to acquire maximal information from realistic experiments we apply specifically fine-tuned Hidden Markov models (HMMs) that consider every data point. Since we cannot give a full introduction to HMM in this article, the reader is referred to the classic introduction by Rabiner[11] or textbooks, e.g. ref. [12]. Herein, we aim to discuss and highlight helpful adaptations of the formalism to smFRET data.

The advantage of these adaptations is demonstrated on the basis of conformational changes of the 90kDa heat-shock protein Hsp90. The presented approach allows us to quantify even the small kinetic effects caused by different ionic strengths, and even different cations. As intra-molecular interactions of proteins depend strongly on the ionic strength of their environment, a systematic salt screen provides valuable information on critical intra-molecular interactions and the origin of conformational stabilization or destabilization. Indeed, we find a clear relationship between the salt-dependence of Hsp90's ATPase activity on the one hand, and its conformational kinetics, on the other hand.

# THEORY

In realistic TIRF experiments, three assumptions apply: (i) no fluorophore saturation, (ii) stochastically independent sources of noise, (iii) negligible laser-independent noise.

The hardware limit for maximal *time resolution* depends on the utilized detector. For EMCCD cameras - still the most frequent detector for single molecule fluorescence time traces - the maximal frame rate is less than 60 Hz. Higher sampling rates are achieved by cropped chip exposure or using alternative detectors, such as s-CMOS cameras or APDs. On the other end of the time window, an extended observation time is desirable for kinetic analysis. This restricts the experimentally applicable laser powers to levels below fluorophore saturation. Consequently, the intensities of excitation and fluorescence scale linearly, and so does the time constant of photo-bleaching (see below). Thus, within the experimentally relevant regime, the following statements in units of time $t$ remain general - independent of the actual sampling rate.

A typical organic fluorophore emits a few million photons before irreversible photo-bleaching[7,8], which defines the *observation time* of an individual molecule. Because these dyes have a high fluorescence quantum yield, the mean of the exponentially distributed total number of excitation/de-excitation cycles is $\widehat{N}_{cycles}^{tot} \approx 10^6$.

In the absence of fluorophore saturation, the time constant of bleaching in units of time $\Delta t$ is given by:

$$\tau_{bl} = \frac{\widehat{N}_{cycles}^{tot}}{n_{cycles}} = \frac{\widehat{N}_{cycles}^{tot}}{\epsilon_{ex} \cdot n_{ph}^{ex}} \quad (2)$$

where $n_{cycles}$ is the number of excitation cycles per $\Delta t$, which is determined by the excitation quantum yield $\epsilon_{ex}$ and the number of excitation photons per $\Delta t$, $n_{ph}^{ex}$. The latter is linked to $P_{Laser}$, the incident laser power at the sample, and the photon energy $h\nu$ by:

$$n_{ph}^{ex} = P_{Laser} \frac{\Delta t}{h\nu} \quad (3)$$

The *SNR* is defined as the mean number of signal photons $n_{ph}^{sig}$ per standard deviation of the noise $\sigma_{ph}^{noise}$:

$$SNR = \frac{n_{ph}^{sig}}{\sigma_{ph}^{noise}} \quad (4)$$

In TIRF experiments, the number of signal photons per $\Delta t$ is defined as the difference between detected photons $n_{ph}^{det}$ and background photons $n_{ph}^{bg}$. It is further determined by $n_{ph}^{ex}$, $\epsilon_{ex}$, the fluorescence quantum yield $\phi_{fl}$ and the detection sensitivity $\delta_{sens}$. Below, the later constants are summarized in $C_{sig}$.

$$n_{ph}^{sig} = n_{ph}^{det} - n_{ph}^{bg} = n_{ph}^{ex} \cdot \epsilon_{ex} \cdot \phi_{fl} \cdot \delta_{sens} \equiv n_{ph}^{ex} \cdot C_{sig} \quad (5)$$

The dominating noise sources (variances) are classified, on the one hand, into laser-dependent noise including i) auto-fluorescence, $n_{ph}^{ex} c_{autofl}$; ii) Raman scattering, $n_{ph}^{ex} c_{scat}$; iii) shot-noise, $n_{ph}^{ex} \cdot C_{sig}$; and on the other hand, laser-independent noise, such as iv) detection noise, $\rho_{det}$, originating from read-out, dark-current and the analog-to-digital converter; or v) additional noise, $\beta$, e.g. from dust. The total noise level of these stochastically independent noise sources is given by the root sum of the respective variances:

$$\sigma_{ph}^{noise} = \sqrt{n_{ph}^{ex} \cdot c_{autofl} + n_{ph}^{ex} \cdot c_{scat} + n_{ph}^{ex} + \rho_{det} + \beta} \qquad (6)$$

Furthermore, it is known from TIRF experiments that laser-dependent noise is by far dominating leading to the approximation:

$$\sigma_{ph}^{noise} \approx \sqrt{n_{ph}^{ex} \cdot c_{autofl} + n_{ph}^{ex} \cdot c_{scat} + n_{ph}^{ex}} = \sqrt{n_{ph}^{ex} \cdot C_{noise}} \qquad (7)$$

With $C_{noise}^2 = c_{autofl} + c_{scat} + C_{sig}$

Equation (4) then becomes:

$$SNR = \sqrt{n_{ph}^{ex} \cdot \frac{C_{sig}}{C_{noise}}} \qquad (8)$$

Finally, combining Equations (2) & (8) results in:

$$\tau_{bl} = \frac{\widehat{N}_{cycles}^{tot}}{\epsilon_{ex} \cdot SNR^2} \cdot \left(\frac{C_{sig}}{C_{noise}}\right)^2 = \frac{const}{SNR^2} \qquad (9)$$

It is worth noting, that by eliminating $n_{ph}^{ex}$, we also got rid of $\Delta t$ and $P_{Laser}$. Consequently, the relation between $\tau_{bl}$ and $SNR$ depends exclusively on fluorophor-specific constants ($\widehat{N}_{cycles}^{tot}$, $\epsilon_{ex}$, $\phi_{fl}$) and setup-specific constants ($\delta_{sens}$, $c_{autofl}$, $c_{scat}$).

**EXPERIMENTAL METHODS**

The *protein constructs* were recombinantly expressed in *E.coli* and purified as previously described[10]. Single cysteines at position 61 or 385 were used for site-specific fluorescent labelling with the FRET donor Atto550 or the acceptor Atto647N, respectively. An artificial c-terminal zipper motif was used to keep dissociated dimers in close proximity. If not stated differently, all chemicals were purchased from Sigma Aldrich.

Single Molecule FRET was measured using a home built TIRF setup as previously detailed[10]. Measurements were performed in potassium buffers (40mM HEPES potassium salt, 10mM $MgCl_2$, KCl as specified, pH 7.5 by HCl) or sodium buffers (40mM HEPES anhydrous, 10mM $MgCl_2$, NaCl as specified, pH 7.5 by NaOH).
Single Molecule data was corrected for background fluorescence, leakage, direct excitation, as well as, dye-specific excitation efficiencies, laser intensities, quantum yields and detector sensitivities, using the 2D stoichiometry vs. efficiency approach[13]. Kinetic models were obtained using the single molecule analysis of complex kinetic sequences (SMACKS)[10].

The ATPase activity of Hsp90 was measured in a regenerative assay[14]: 0.2 mM NADH, 2 mM phosphoenol pyruvate (PEP), 2 U/ml pyruvate kinase (Roche), 10 U/ml lactate dehydrogenase (Roche) coupled to NADH oxidation, which was followed as a decrease in absorption at 340nm. All-sodium or all-potassium conditions were prepared using corresponding reagents: NADH Di-Na (Roche) or Di-K, PEP Na or K (Bachem), ATP Mg-salt. The above enzymes were further dialyzed against the corresponding low salt buffer: 40mM Hepes, 10mM $MgCl_2$, 50mM NaCl or KCl, pH 7.5 by NaOH or KOH, respectively. Measurements were performed at 37°C in 40mM Hepes, the indicated NaCl or KCl concentration, 10mM $MgCl_2$, pH 7.5. Each measurement was followed by radicicol inhibition and addition of excess ADP as a positive control for regeneration.

# RESULTS & DISCUSSION

## A. The input data

Figures 2a shows an illustration of the TIRF experiment on an Hsp90 dimer, where one monomer is labelled at residue 61 with Atto550 and the other one at residue 385 with Atto647N (see Experimental Methods for details). Figures 2b shows the resulting experimental raw data for three different potassium concentrations. The individual traces recorded under 50, 150 or 750mM KCl show no significant difference: mainly low FRET is observed - representing v-shaped, open conformations of Hsp90 - with intermittent high-FRET spikes of varied length, representing sporadic closure. Interestingly, these dynamics are observed in the absence of an external energy source other than thermal energy.

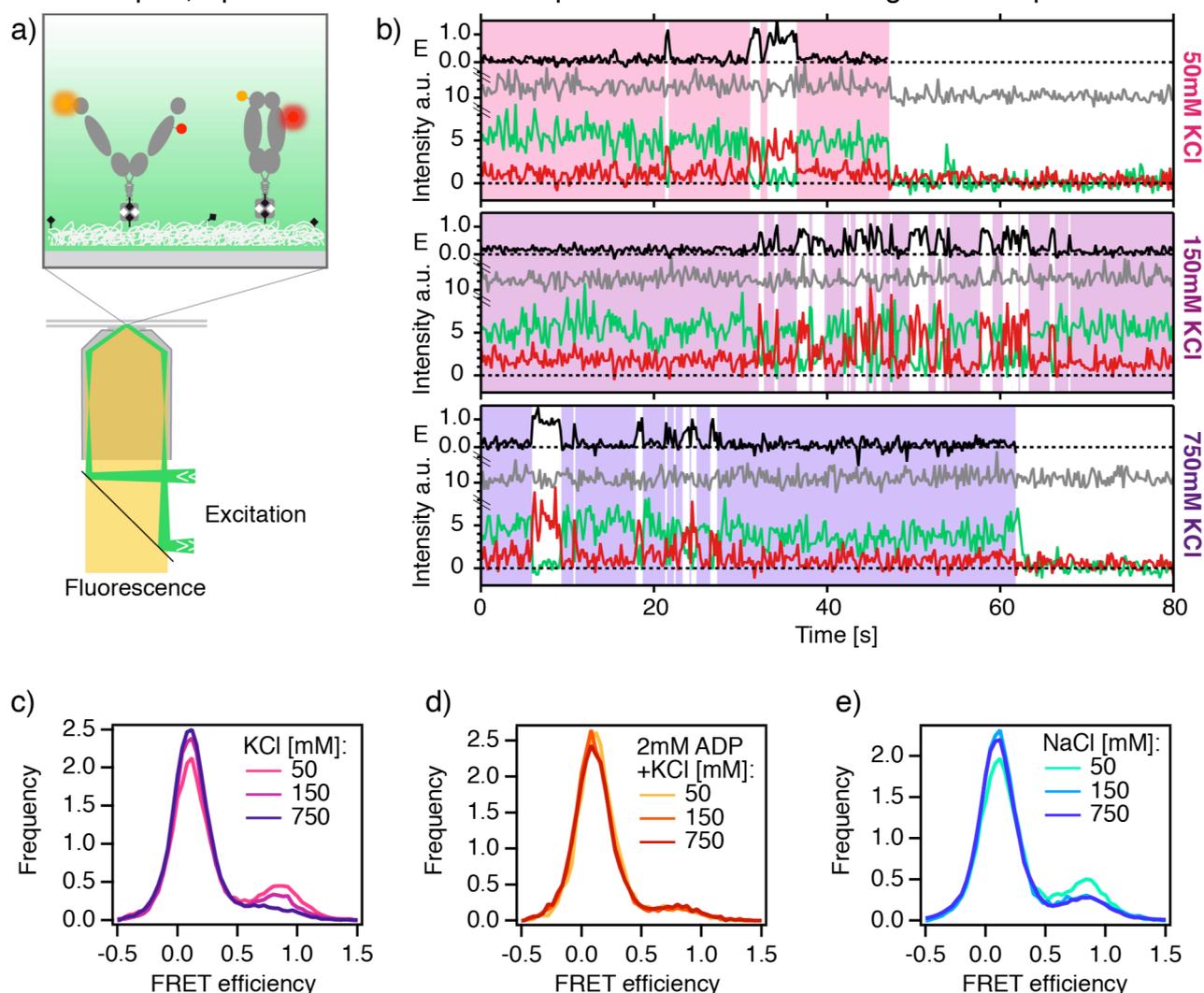

**Figure 2:** (a) Illustration of the TIRF experiment. An objective-type TIRF microscope was used to record the fluorescence of surface-immobilized protein molecules. FRET between two specifically attached dyes allows one to distinguish v-shaped, N-terminally *open* (left) from *closed* (right) conformations. (b) Experimental raw data recorded in real time under three different potassium concentrations as indicated. Fluorescence intensities are color coded: donor (green), FRET sensitized acceptor (red), directly excited acceptor (gray) and FRET efficiency, E (black). Colored or white overlays indicate HMM-derived low-FRET or high-FRET states, respectively. (c-e) Cation dependence of Hsp90's conformations for potassium chloride (c), including ADP (d), or sodium chloride (e) as specified.

Accordingly, the FRET efficiency histograms of many such traces show a large low-FRET population and only a small high-FRET population. While the peak positions - representing the predominant conformations - remain unaffected, a slight but reproducible depopulation of Hsp90's closed conformations is observed under increasing KCl concentrations (Figure 2c). Interestingly, this effect was abolished by the addition of ADP (Figure 2d). For similar sodium concentrations (Figure 2e), the trend towards prevalence of open conformations was less pronounced.

The equilibrium shift towards open conformations under high salt conditions can either be caused by screening of electrostatic cross-protomer interactions, or by the stabilization of hydrophobic interactions within the open conformation, thereby burying potential hydrophobic cross-protomer contacts. Interestingly, the more pronounced effect observed for potassium as compared to sodium ions matches the Hofmeister trend, i.e. stronger hydrophobic interaction under the larger cation with lower charge density[15]. This hints towards a dominating role of hydrophobic stabilization of the open conformation. It is further in line with cation-induced Hofmeister effects reported for cytochrome c[16].

Further kinetic insight is gained using SMACKS. It comes in two steps: the first one applies trace-wise HMMs to capture the heterogeneity between individual molecules. This serves as a basis for the second step consisting of a semi-ensemble HMM optimization, which directly provides one global kinetic rate model for the entire data set.

### B. Step I: trace-by-trace HMM

The above data - the individual traces and the aggregate histograms - suggests that a 2-state model is a good starting point for further kinetic analysis using SMACKS. Thus, we use a 2-state hidden Markov model $\lambda(\pi, A, B)$, which is parameterized by start ($\pi$), transition ($A$) and emission ($B$) probabilities in a first trace-wise optimization.

Two adaptations are useful when dealing with smFRET data: first *fluorescence* time traces - not *FRET efficiency* traces - are the preferred input data. The robustness of the HMM with respect to uncorrelated noise is significantly increased by exploiting the original two observables - donor and acceptor fluorescence - instead of the FRET efficiency (only one observable). In addition, FRET efficiencies come with unfavorable spikes - due to occasional, noise related division by zero - which are absent in the original fluorescence traces. Therefore, no previous smoothing is required if fluorescence traces are used as HMM input. Figure 3a demonstrates both, the superiority of the 2D approach regarding uncorrelated spikes and also noise induced poles of the FRET efficiency.

Corresponding 1D and 2D histograms are displayed in Figure 3d,e, further highlighting the higher amount of information in the 2D case. Such fluorescence signals are appropriately described by a 2D Gaussian probability density function (PDF) for each state. These are parameterized by the vector of means, $\mu_i$, for each state $i$, and the covariance matrix, $V_i$, of the donor and acceptor intensities. A representative emission PDF is displayed in Figure 4a (top right).

As a second adaptation to FRET data, we exploit the physical relation between the means of the donor and acceptor intensities, $\mu_{i,A}$ and $\mu_{i,D}$, which must add up to the average total intensity of the respective trace:

$$\langle I_{tot} \rangle = \sum_{t=1}^{T} \frac{(x_{t,A} + x_{t,D})}{T} = \mu_{i,A} + \mu_{i,D} = const. \;\; \forall i \tag{10}$$

where $x_{t,A}$ and $x_{t,D}$ are the acceptor and donor intensities at time $t$. And $T$ denotes the total time of a single trace $i$. As a result, the available parameter space for the means, $\mu_{i,A} + \mu_{i,D}$, shrinks to one line. This FRET line is displayed in Figure 4a(top, right) in red. To comply with experimental variations between individual molecules as seen in Figure 3e, the line is determined individually for each molecule.

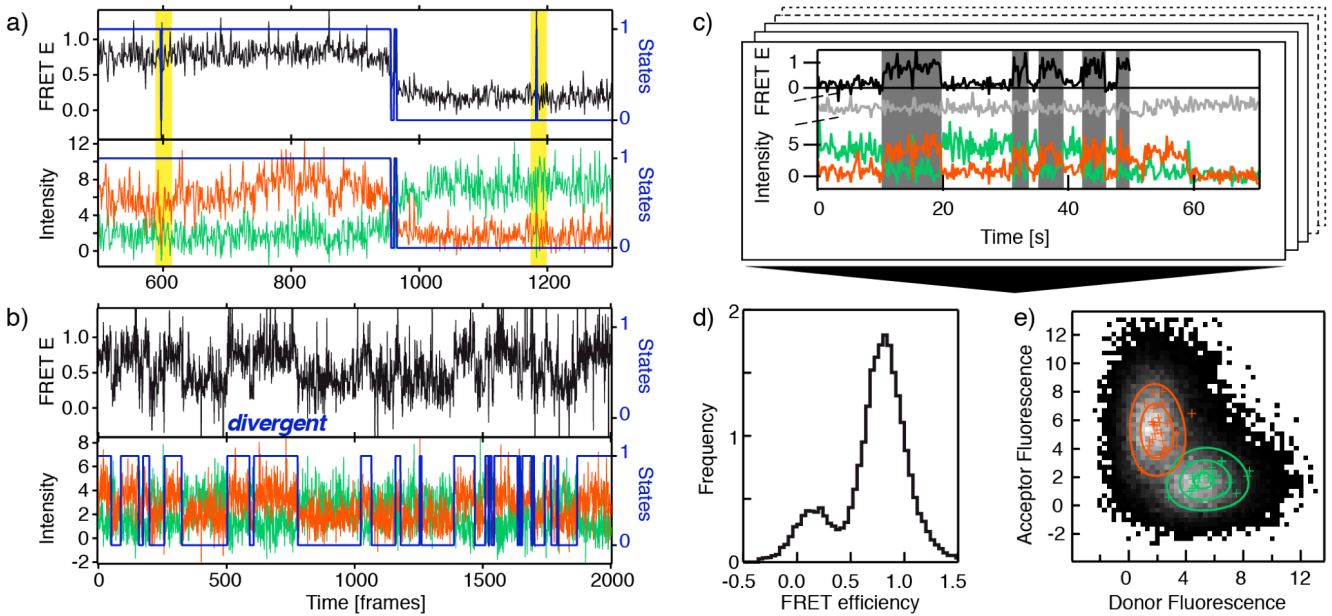

**Figure 3:** (a, b) The superior robustness of 2D HMM demonstrated by two example traces: FRET efficiency (FRET E, black), fluorescence intensity of the donor (green), acceptor (orange), Viterbi path (blue, right axis, state 0: low FRET, state 1: high FRET). (a) Blink events (highlighted in yellow) are misinterpreted by FRET efficiency based 1D HMM. (b) 1D HMM diverges under high noise conditions. Ergo, the Viterbi path is not defined. In contrast, 2D HMM still derives a suitable Viterbi path. (c) smFRET data as input for HMM: FRET E (black), intensity of the donor (green), acceptor (orange), directly excited acceptor (gray), Viterbi path as gray and white overlays indicating high- and low-FRET states, respectively. (d) The FRET efficiency histogram of multiple traces provides only 1D information, although 2D information was originally recorded. (e) 2D fluorescence histogram of smFRET data (black to light gray: minimal to maximal counts, white: no counts). The means of 36 individual donor and acceptor traces are indicated with green and orange markers, respectively. Global Gaussians, as derived for the entire data set, are displayed as corresponding contours.

Each model $\lambda(\pi, A, B)$ is iteratively rated by the forward-backward algorithm and optimized by the Baum-Welch algorithm until convergence to maximum likelihood. Because the likelihood function reaches very flat plateaus between steep descents, it is not a convenient reporter for convergence. By contrast, the normalized changes of the diagonal entries of the transition matrix have proven useful for monitoring convergence of the HMM:

$$\text{Normalized Changes} = \sum_{i=0}^{n-1} \frac{|a_{ii} - a'_{ii}|}{a_{ii}} \tag{11}$$

where $a'_{ii}$ are the diagonal matrix elements of the previous iteration and the sum goes over all states $i$. In this work, no further changes were found, once this quantity fell below $10^{-8}$.

Next, the Viterbi algorithm is used to compute the most probable state sequence for each trace given the trace-specific model. The visual comparison of the resulting Viterbi path to the original input data serves as a quality control of the underlying parameters. Quite conveniently, HMM emulates a characteristic requirement for single-molecule fluorescence data, by searching for flat plateaus. Therefore, traces that are not well described by the Viterbi path, are often sub-quality traces and as such sorted out. On the other hand, the apparent model must be revised if the HMMs fail repeatedly at good quality traces (with respect to signal-to-noise, signal regularity etc.).

As a direct consequence of the finite observation time (due to bleaching) not every time trace shows transitions between distinct FRET efficiencies. Importantly, even such *static* traces contain kinetic information. We include static traces using the mean emission PDFs of the remaining data set, because they would not converge sensibly in a trace-by-trace run with more than one state. Typical Hsp90 data sets contained about 30% static traces.

**C. Step II: semi-ensemble HMM**

In a second step a semi-ensemble HMM run is performed to derive one kinetic model based on a set of traces. To this end, the global start and transition probabilities are optimized, while the individual emission PDFs, trained in the last step, are held fixed (see illustration in Figure 4a).

Distinct kinetic states that cause experimentally indistinguishable smFRET signals are frequently observed with proteins[10,17-19]. Using semi-ensemble HMM, such kinetic heterogeneity can be investigated by comparing the fit of different state models including duplicates and triplicates of the apparent states. It is obvious that with an increasing number of degrees of freedom also the likelihood of the model increases. In the extreme case, a model could consist of one state per time step and thus describe the data perfectly - but without any physical meaning. Therefore, parsimony criteria are commonly used to identify the optimal model - that is to say, a model that describes the data well, while keeping the model complexity moderate. Here the Bayesian information criteri-on (BIC)[20] is used for model selection, similar to earlier studies[21-24]. It balances the likelihood, $\mathcal{L}$, against the number of free parameters, k, and the number of data points, n:

$$BIC = -2 \cdot \ln(\mathcal{L}) + k \cdot \ln(n) \qquad (12)$$

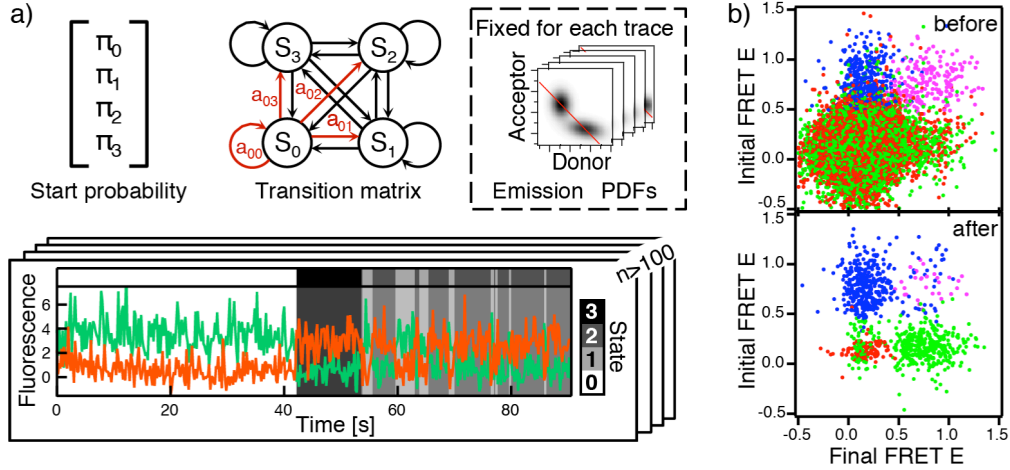

**Figure 4:** (a) Semi-ensemble HMM optimizes a global kinetic model based on a complete data set (normally >100 traces). While the kinetic parameters - start probabilities and transition matrix - are optimized globally, the predetermined, individual emission PDFs are held fixed. This allows further to identify states not only by a characteristic signal, but also based on their kinetic behavior. For the example trace displayed, this results in a Viterbi path (overlays) with 4 kinetic states despite only 2 distinguishable FRET efficiencies. (b) Transition maps "before" and "after" optimization of the HMM. A fitting rate model generates well defined clustering: the mean FRET efficiencies of the dwell preceding a transition (initial FRET E) are plotted against those of the following dwell (Final FRET E). The initial state is color-coded: state 0 (red), 1 (green), 2 (blue), 3 (pink), for further details see Ref. [10].

Once the optimal number of states is deduced and the model fits the raw data as shown by the transition map (Figure 4b), we use the procedure by Bruno et al.[25] to find the simplest, plausible reaction scheme given the data. Based on the 4-state model with 2 open (o) and 2 closed (c) states ($N_o = N_c = 2$ as previously determined by BIC) we compare models of the so-called canonical "MIR"-form (manifest interconductance rank) of rank 1 (linear o-o-c-c) and rank 2 (cyclic -o-o-c-c-) in a likelihood ratio test (LR):

$$\text{LR} = 2 \cdot [\ln(\mathcal{L}_{R2}) - \ln(\mathcal{L}_{R1})] \begin{cases} \leq \chi^2_{0.95, df=2} & \Rightarrow \text{rank 1} \\ > \chi^2_{0.95, df=2} & \Rightarrow \text{rank 2} \end{cases} \qquad (13)$$

where $\mathcal{L}_{Rx}$ denotes the likelihood of rank $x$. The null hypothesis (rank 1 model) is rejected if the likelihood ratio exceeds the 95% confidence interval given by the $\chi^2$-distribution for 2 degrees of freedom ($df$). Note that one missing link equals a difference of two transitions. For apo Hsp90, rank 1 was found.

Normally, the next step is the determination of the number of links $N_l$ within this rank $R$ by comparing different schemes by BIC. The number of mathematically identifiable links is limited to $N_l \leq R(N_o + N_c - R)$. So for the discussed case ($N_o = N_c = 2; R = 1$) the only option is $N_l = 3$ and the link determination is redundant. Please note that models with the same rank and the same number of links are mathematically equivalent. Thus, without prior knowledge or further experimental data, we cannot discriminate the models displayed in Figure 5a from the kinetic data alone. Further information on the interpretation of degenerate state models is given in [25-27]. Regarding our apo Hsp90 data, the existing structural information supports the linear o-o-c-c model, Figure 5a (top).

A model without an uncertainty estimate is worthless. Thus we calculate confidence intervals similar to refs. [10,22]. As illustrated in Figure 5b, every rate is gradually moved away from its maximum likelihood estimator. At every step the likelihood is evaluated and compared to the original maximum likelihood in a likelihood ratio test. As the likelihood ratio follows a $\chi^2$-distribution, the 95% confidence bounds are reached where the likelihood ratio crosses $\chi^2_{df=1}(\alpha = 0.95) = 3.84$.

In addition, multiple runs with random start parameters are performed to recognize potential local minima. The analysis of subsets of the data is useful to estimate the data-set heterogeneity. Finally, it is illustrative to re-simulate dwell-time distributions given the obtained model. Comparison to the original, experimental dwell-time distributions gives a qualitative estimate of the fit of the model and the data. If the experimental bleach rate and data-set size are retained, it reveals also the purely statistical variability of the results.

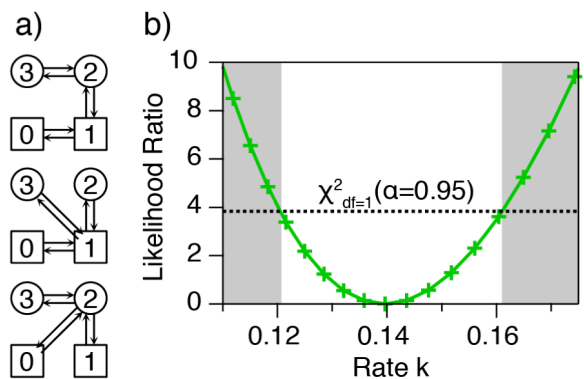

**Figure 5:** (a) Three models that are mathematically equivalent as they all have 2+2 states and rank 1. (b) Determination of confidence bounds: every rate k is gradually moved away from its maximum likelihood estimator. The likelihood ratio between the old and new model is displayed as a function of the modified rate constant. The 95% confidence bounds are reached where the likelihood ratio crosses $\chi^2_{df=1}(\alpha = 0.95) = 3.84$.

## D. Salt effects on conformational & functional kinetics of Hsp90

Figures 6a-d show Hsp90's transition rates between open and closed conformations and their uncertainties deduced by SMACKS, as described above. In agreement with earlier findings, SMACKS infers a kinetic 4-state model with 2 closed and 2 open states. The 2 closed states presumably differ in a small secondary structure element - the very N-terminal $\beta$-sheet providing extra stabilization to the long-lived closed state by reaching over to the opposite monomer, whereas a shorter-lived closed state is found in the absence of those additional cross-monomer contacts[28-30]. Conversely, the 2 open states mimic the kinetics of a much larger conformational ensemble, which is beyond the time and distance resolution of smFRET time traces, but detectable using the large micro-time statistics of confocal smFRET in solution[31].

Figure 6a shows that in addition to the observed population shift, there is an overall growth of the rate constants with increasing potassium concentrations. Conversely, no significant change is observed under equal sodium concentrations (Figure 6b). Although all shifts are small, they occurred consistently upon buffer change: e.g. the data at 150mM KCl was measured after that at 750mM KCl. It agrees well with a previous data set under the same conditions ("prev. 150" in Figure 6a).
So the energy barriers between Hsp90's conformations shrink gradually under increasing KCl concentrations, which can be interpreted as an increased overall flexibility. Along with that, the mean number of transitions per trace grows for increasing potassium - but not sodium - concentrations (cf. Figure 6d).

Despite the significantly different effects of potassium and sodium on the transitions, the effects on the ATPase activity are similar. Fig. 6e shows that Hsp90's ATPase activity grows about 3-fold with increasing potassium or sodium concentrations between 50mM and 1M, which was consistently found for three individual Hsp90 constructs. A similar effect of potassium chloride on the ATPase activity has also been measured before[32].

Up to now, an increased ATPase rate was typically related to an increased population of closed conformations in Hsp90 (e.g by mutations or cochaperones), which was commonly interpreted as Hsp90's active state[29]. This is clearly not the case here, where the occupation of the closed state even decreases with increasing salt (Fig. 2c,e). Therefore, in the following we discuss potential other mechanisms. Essentially two things can increase the ATPase rate: either a more efficient hydrolysis process or altered nucleotide affinities. A faster product release (i.e. lower ADP affinity) could increase the ATPase rate. This is however unlikely due to the observed repression of the salt effect in the presence of ADP (Fig. 2d). But an increased ATP affinity cannot be ruled out, because salt-dependent ATP affinities are very difficult to determine precisely, both by fluorescence related methods (due to artifacts of the nucleotide-dye conjugate) and by calorimetry (due to hydrolysis induced heat). On the other hand, local changes in the ATP binding pocket can also lead to a more efficient hydrolysis process. Such local changes are not yet accessible by smFRET, but they are currently investigated by MD simulations. Altogether, we guess that local rearrangements at the nucleotide binding pocket are responsible for the observed increased ATPase rate.

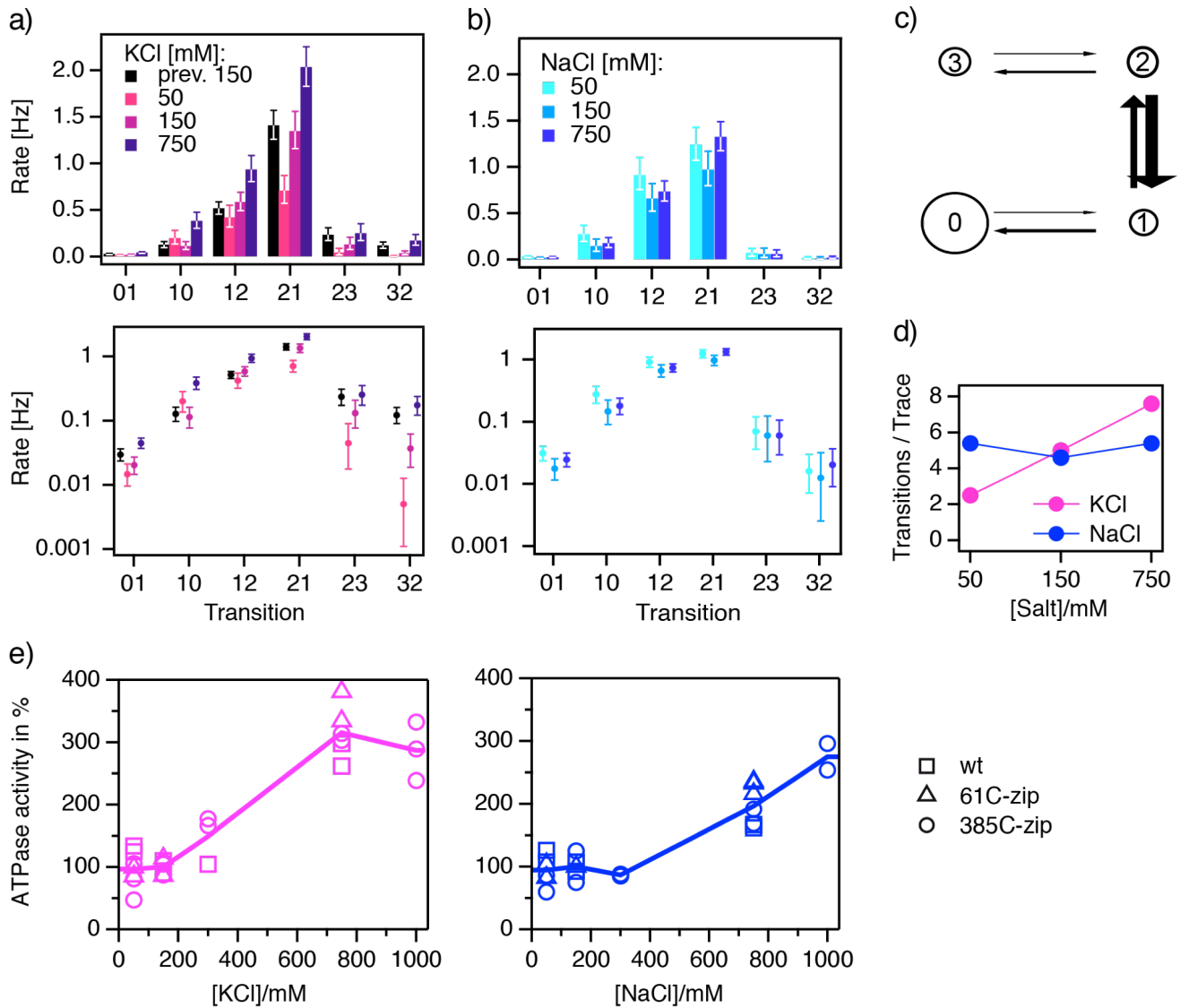

**Figure 6:** Dissimilar cation dependence of Hsp90's kinetics under varied potassium (a) or sodium (b) concentrations. The rates are labeled according to the state model in (c). State 0,1: low FRET; state 2,3: high FRET; circle sizes represent populations; arrow weights represent transition rate constants. (d) The average number of transitions observed per trace for each data set. (a-d) The number of molecules included in the dataset prev. 150 / 50 / 150 / 750mM KCl is 154 / 107 / 102 / 129, respectively; and for 50 / 150 / 750mM NaCl it is 105 / 70 / 140, respectively. (e) Hsp90's ATPase activity under varied cation conditions: KCl, left; NaCl, right. Measurements were performed with 3 different Hsp90 variants: wild-type (wt), 61C with C-terminal zipper (zip) and 385C with C-terminal zipper. Individual rates were normalized to the value obtained using 150mM monovalent cation. For KCl these values were: 0.8 / 0.8 / 0.7 ATP/min/monomer; for NaCl: 1.2 / 0.7 / 0.4 ATP/min/monomer (in the above order). The lines connect the average measured rates.

**CONCLUSION**

In this study we detail the HMM-based single molecule analysis of complex kinetic sequences (SMACKS), and use this tool to determine the salt-dependent conformational kinetics of the protein Hsp90. We find that all rate constants for Hsp90's large conformational changes become faster with increasing KCl concentrations, but not NaCl concentrations. This implies lower energy barriers between individual conformations in the presence of KCl. In addition, high salt conditions shift the conformational equilibrium towards open conformations. In agreement with the Hofmeister trend, a larger effect was detected with KCl. This suggests that the observed shift to open conformations is caused by strengthened hydrophobic intra-monomer interactions, rather than weakened electrostatic interactions.

At the same time an increase of Hsp90's ATPase activity with growing salt concentrations was found. This is remarkable, as the closed conformation is generally accepted to be Hsp90's active state. Given the conservative effect of ADP, an increased product release due to a lower affinity for ADP seems unlikely. However, further experiments are needed to eventually pin down if the cause is an altered catalysis process, or a salt-dependent nucleotide affinity.

The strength of our approach lies in the combination of robustness regarding experimental noise, on the one hand, and the accumulation of fragmented kinetic information in one global model, on the other hand. It is this shortcut from raw data to kinetic information - notably avoiding the earlier detour over biased dwell times - that makes SMACKS particularly efficient in distilling information from experimental data. Herein, we focused primarily on adaptations of the HMM formalism to smFRET, but SMACKS is suitable for all kinds of single molecule data.

**ACKNOWLEDGMENTS**
We thank Jens Timmer, Markus Götz and Philipp Wortmann for helpful discussions and the latter two for protein donations. We thank Jolanta Vorreiter for wet lab assistance.
This work was funded by the European Research Council through ERC grant agreement no. 681891.